\documentclass[paper,prb]{revtex4}%
\usepackage{amsfonts}
\usepackage{amsmath}
\usepackage{amssymb}
\usepackage{graphicx}
\usepackage{amscd}
\usepackage{acronym}%
\begin{document}
\title[Lattice superconductor]{London's limit for the lattice superconductor}
\author{S.A. Ktitorov}
\affiliation{A.F. Ioffe Physico-Technical Institute of the Russian Academy of Sciences,
Polytechnicheskaja str. 26, St. Petersburg 194021, Russia and Department of
Physics, St. Petersburg. State Electroengineering University, Prof. Popov str.
5, St. Petersburg 197376, Russia.}

\begin{abstract}
A stability problem for the current state of the strong coupling
superconductor has been considered within the lattice Ginzburg-Landau model.
The critical current problem for a thin superconductor film is solved within
the London limit taking into account the crystal lattice symmetry. The current
dependence on the order parameter modulus is computed for the superconductor
film for various coupling parameter magnitudes. The field penetration problem
is shown to be described in this case by the one-dimensional sine-Gordon
equation. The field distribution around the vortex is described at the same
time by the two-dimensional elliptic sine-Gordon equation.

\end{abstract}
\maketitle








\section{Introduction}

\qquad An influence of the external magnetic field on the superconducting
state near the upper critical field is intensively studied during\ the last
decade . It was observed \cite{brezin} that a strong magnetic field near the
upper critical field point makes superconducting order parameter fluctuations
quasionedimensional and, therefore, strong. Nonrenormalizability of the theory
at a strong field leads to impossibility to apply the renormalization group
technique. The lattice translation symmetry is understood to be crucial for
the renormalizability of the theory \cite{We}. The Ginzburg - Landau effective
action is shown to acquire the Harper's operator form of the kinetic term that
restores a three-dimensional nature of fluctuations and renormalizability of
the theory. It is important to consider, what physical consequences follow
from this form of the kinetic term in the Ginzburg-Landau theory. However, a
sophisticated mathematical structure of this operator makes working with it
rather difficult. Here we consider an alternative approach to the problem
using the quasiclassical approach in the London's limit of extremely small
coherence length.

\section{Equations of electrodynamics for the narrow-band superconductor}

We start from the lattice Ginzburg-Landau functional which was derived in
\cite{sherst} and used for a study of the critical behaviour in \cite{We}:%

\begin{equation}
F=\int d^{3}x\left[  \psi^{\ast}(\mathbf{x)}\epsilon(-i\hslash\nabla-\frac
{2e}{c}\mathbf{A(x)})\psi(\mathbf{x})+\tau\psi^{\ast}(\mathbf{x)}%
\psi(\mathbf{x})+\frac{g}{2}(\psi^{\ast}(\mathbf{x)}\psi(\mathbf{x}%
))^{2}+\frac{\mathbf{B}^{2}}{8\pi}\right]  , \label{GLF}%
\end{equation}

where $\mathbf{B=}rot\mathbf{A}$ is the magnetic induction, , $\tau
=\alpha\left(  T-T_{c}\right)  /T_{c},$ $T_{c}$ is the mean-field
approximation superconducting transition temperature, $\epsilon(\mathbf{p})$
is a 3d-perodic function of the quasimomentum with periods $2\pi/a_{1},$
$2\pi/a_{2},$ $2\pi/a_{3}$. Operators $\nabla$ and $\mathbf{A(x)}$ commute if
we choose the Coulomb-London gauge $div\mathbf{A}=0$. This functional can be
rewritten in the Wannier
\begin{equation}
F=\sum_{\mathbf{m,m}^{\prime}}J_{\mathbf{m-m}^{\prime}}\exp\left[  i\frac
{2e}{\hbar c}\int_{\mathbf{m}}^{\mathbf{m}^{\prime}}d\mathbf{l\cdot
A(x)}\right]  \phi_{\mathbf{m}^{\prime}}^{\ast}\phi_{\mathbf{m}}%
+\sum_{\mathbf{m}}\left[  \tau\phi_{\mathbf{m}}^{\ast}\phi_{\mathbf{m}%
}+g\left(  \phi_{\mathbf{m}}^{\ast}\phi_{\mathbf{m}}\right)  ^{2}\right]
+\int d^{3}x\frac{\mathbf{B}^{2}}{8\pi} \label{wannier}%
\end{equation}
or Bloch
\begin{align}
F  &  =\sum_{\mathbf{q}}\left[  \varphi_{\mathbf{q}}^{\ast}\epsilon
(\mathbf{q}-\frac{2e}{\hbar c}\mathbf{A(i}\frac{\partial}{\partial\mathbf{q}%
}\mathbf{)})\varphi_{\mathbf{q}}+\tau\varphi_{\mathbf{q}}^{\ast}%
\varphi_{\mathbf{q}}\right]  +\nonumber\\
&  \sum_{\mathbf{q}_{1}+\mathbf{q}_{2}+\mathbf{q}_{3}+\mathbf{q}_{4}=0}%
\frac{g}{2}\varphi_{\mathbf{q}_{1}}^{\ast}\varphi_{\mathbf{q}_{2}}%
\varphi_{\mathbf{q}_{3}}^{\ast}\varphi_{\mathbf{q}_{4}}+\int d^{3}%
x\frac{\mathbf{B}^{2}}{8\pi} \label{Bloch}%
\end{align}
representations. The lattice vectors $\mathbf{m}$ numerate the lattice cites
in the plane perpendicular to the magnetic field.

Varying Eq. (\ref{GLF}) over the vector potential $\mathbf{A(x)}$, we obtain
the Maxwell equation
\begin{equation}
\frac{c}{4\pi}rot\mathbf{B=j} \label{max}%
\end{equation}
with the following current density%

\begin{equation}
\mathbf{j=}\frac{e}{2}\left[  \psi^{\ast}(\mathbf{x)v}(-i\hslash\nabla
-\frac{2e}{c}\mathbf{A(x)})\psi(\mathbf{x})+\psi(\mathbf{x)v}(i\hslash
\nabla-\frac{2e}{c}\mathbf{A(x)})\psi^{\ast}(\mathbf{x})\right]  ,
\label{current}%
\end{equation}
where $\mathbf{v(p)=}\partial\mathbf{\epsilon}/\partial\mathbf{p}$ is the
group velocity of the order parameter wave packet. Let us present the complex
order parameter $\psi(\mathbf{x})$ in the form%

\begin{equation}
\psi(\mathbf{x})=R(\mathbf{x})\exp\left(  i\phi(\mathbf{x}\right)  ,
\label{order}%
\end{equation}
where $R^{2}=n_{s}$ is the density of superconducting electrons. When the
coherence length is small in comparison with the field penetration length,
$\tau$ is large that is equivalent to a presence of the small parameter in the
kinetic term. Then the order parameter modulus $R$ can be taken as a spatially
uniform solution of Eq. (\ref{GLF}):%

\begin{equation}
\tau(T)R_{0}=-gR_{0}^{3} \label{modulus}%
\end{equation}
almost everywhere. A small parameter before the higher derivative does not
guarantee here a complete spatial uniformity: this is just a loophole for the
vortex core solution to appear. Assuming the order parameter modulus to be
costant and the phase gradient to be slow we come to the expression for the
current in the London limit:%

\begin{equation}
\mathbf{j=}en_{s}^{0}\mathbf{v}(\nabla\phi-\frac{2e}{\hbar c}\mathbf{A(x)}),
\label{current1}%
\end{equation}
where $n_{s}^{0}=R_{0}^{2}$. Introducing the gauge invariant vector field%

\begin{equation}
\mathcal{A=}\mathbf{A}-\frac{\hbar c}{2e}\nabla\phi, \label{gaugeinv}%
\end{equation}
we can rewrite the Maxwell equation Eq. (\ref{max}) in the form%

\begin{equation}
\frac{c}{4\pi}rotrot\mathcal{A}=-eR_{0}^{2}\mathbf{v}(\frac{2e}{\hbar
c}\mathcal{A}). \label{basic}%
\end{equation}

Eq. (\ref{basic}) can be easily reduced to the standard London equation%

\begin{equation}
\delta^{2}\nabla^{2}\mathbf{A}-\mathbf{A=}0 \label{london}%
\end{equation}

with $\delta^{2}=mc^{2}/\left(  4\pi e^{2}R^{2}\right)  $ in the continuum
approximation limit $\mathbf{v(p)=p/}m$. Eq. (\ref{basic}) will be applied
below to some classical problems of superconductivity in order to see, what
physical consequences follow from an explicit taking into account the crystal
translation symmetry.

\section{Critical current in the thin film}

\bigskip

We analize the case of the simple tetragonal lattice with the $\epsilon
(\mathbf{p})$ function of the form%

\begin{equation}
\epsilon(\mathbf{p})=\Delta_{\bot}(2-\cos\frac{p_{x}a}{\hbar}-\cos\frac
{p_{y}a}{\hbar})+\Delta_{\Vert}(1-\cos\frac{p_{z}b}{\hbar}). \label{spectrum}%
\end{equation}

Let us consider the film of thickness $d$ with the current $\mathbf{j}$
parallel to the $x$-axis. We assume $d<<\xi(T),d<<\delta(T)$, where $\xi(T)$
is the coherence length that secure uniformity across the film thickness
respectively of the order parameter $R$ and the current density $\mathbf{j.}$
Using Eqs. (\ref{GLF)}, (\ref{order}), (\ref{current1}) and (\ref{gaugeinv}),
we can write the following free energy minimum conditions (neglecting the
magnetic field effect as it is usually done in such problems)%

\begin{equation}
j_{x}\mathbf{=-}eR^{2}v_{x}(2e/\left(  \hbar c\right)  \mathcal{A}),
\label{current2}%
\end{equation}

\begin{equation}
\epsilon_{x}(2e/\left(  \hbar c\right)  \mathcal{A})R+\tau R+gR^{3}=0.
\label{gl}%
\end{equation}

Here
\begin{align}
\epsilon_{x}(q_{x})  &  =\Delta_{\bot}(1-\cos q_{x}a),\label{band}\\
v_{x}(q_{x})  &  =\frac{\partial\epsilon_{x}(q_{x})}{\hbar\partial q_{x}%
}=a\Delta_{\bot}/\left(  \hbar\right)  \sin q_{x}a. \label{velocity}%
\end{align}

Eliminating $v_{x}$ and $\epsilon_{x}$ from Eqs. (\ref{band}) and
(\ref{velocity)} with a use of the relation%

\begin{equation}
\epsilon_{x}=\Delta_{\bot}\left[  1-\sqrt{1-\left(  \frac{\hbar v_{x}}%
{a\Delta_{\bot}}\right)  ^{2}}\right]  \label{epsilon-v}%
\end{equation}

we obtain the following relation between $j_{x}$ and $R:$%

\begin{equation}
\Delta_{\bot}R\left(  1-\sqrt{1-\left(  \hbar/\left(  ea\Delta_{\bot}\right)
\right)  ^{2}j_{x}^{2}/R^{4}}\right)  -\left\vert \tau\right\vert R+gR^{3}=0.
\label{j-R0}%
\end{equation}

Introducing dimensionless variables $J$ and $f$ according to formulae%

\begin{align}
j_{x}  &  =ea\Delta_{\bot}R_{0}^{2}J/\hbar,\label{J}\\
R  &  =fR_{0}, \label{f}%
\end{align}

where $R_{0}^{2}=\left\vert \tau\right\vert /g$ is a non-trivial solution of
eq. (\ref{modulus}), we arrive at the following result%

\begin{equation}
1-\sqrt{1-J^{2}/f^{4}}=k\left(  1-f^{2}\right)  , \label{j-R}%
\end{equation}

where $k=\left\vert \tau\right\vert /\Delta_{\bot}.$ In the limit of $\hbar
v_{x}/\left(  a\Delta_{\bot}\right)  \rightarrow0$ and, therefore,
$\lambda\rightarrow0,$ Eq. (\ref{j-R}) reduces to the classical form for the
continuum approximation \cite{degennes}:%

\begin{equation}
J^{2}=2kf^{4}(1-f^{2}) \label{degennes}%
\end{equation}

Two plots for the dimensionless current $J$ dependence on the dimensionless
order parameter $f$ are presented in fig. 1 for the cases of $k\rightarrow0$
and $k=1.$ A dependence of the maximum position $f_{m}$ on $k$ is depicted in
fig. 2.

\bigskip

\section{Field penetration problem}

Let us return to Eq. (\ref{basic}). It can be used in order to determine the
vortex structure outside the core. For the case of the kinetic term in the
form of Eq. (\ref{spectrum}) we can express Eq. (\ref{basic}) as following%
\begin{equation}
\frac{c}{4\pi}rotrot\mathcal{A}=-eR_{0}^{2}\mathbf{v}(e\mathcal{A/}\left(
\hbar c\right)  ), \label{field}%
\end{equation}
or, in London's gauge $div\mathcal{A}=0$:%

\begin{equation}
\frac{c}{4\pi}\nabla^{2}\mathcal{A}-eR_{0}^{2}\mathbf{v}(\mathcal{A}2e/\left(
\hbar c\right)  )=0. \label{laplace}%
\end{equation}
Assuming the dispersion law Eq. (\ref{spectrum}) to be valid, we can write Eq.
(\ref{laplace}) for the $x$-component:%

\begin{equation}
\nabla^{2}\mathcal{A}_{x}-4\pi eR_{0}^{2}a\Delta/\left(  \hbar c\right)
\sin(2\pi\mathcal{A}_{x}a/\Phi_{0})=0, \label{compon}%
\end{equation}

where $\Phi_{0}{\large =}2\pi\hbar c/\left(  2e\right)  $ is the Onsager flux
quantum. This equation can be reduced for the weak field case to
(\ref{london}). It is convenient to introduce the variable%

\begin{equation}
2\pi\mathcal{A}_{x}a/\Phi_{0}=\chi. \label{chi}%
\end{equation}

Then Eq. (\ref{compon}) can be rewritten in the form%

\begin{equation}
\nabla^{2}\chi-\frac{1}{\delta^{2}}\sin\chi=0. \label{sine}%
\end{equation}

\bigskip

Let us consider the classical one-dimensional field penetration problem basing
on Eq. (\ref{compon}). In the one-dimensional case it reads%
\begin{equation}
\frac{d^{2}\chi}{dx^{2}}-\frac{1}{\delta^{2}}\sin\chi=0. \label{1d}%
\end{equation}

Eq. (\ref{1d}) is the well-known equation for the pendulum (but with different
sign); the one-dimensional penetration is described by the solution
\cite{soliton}:%

\begin{equation}
\chi=\arcsin\left\{  \pm cn\left[  -\frac{x-x_{0}}{\kappa\delta}%
;\kappa\right]  \right\}  , \label{penetration}%
\end{equation}

where $cn(x)$ is the elliptic cosine; $\kappa$ is the elliptic module. Notice
that the field penetration problem for the lattice model of a superconductor
has a similarity with the fluxon problem in the one-dimensional Josephson contact.

\section{Vortex solutions}

The field distribution around vortices is described by Eq. (\ref{sine}), which
is better to rewrite in the form%

\begin{equation}
\frac{\partial^{2}\chi}{\partial x^{2}}+\frac{\partial^{2}\chi}{\partial
y^{2}}=\frac{1}{\delta^{2}}\sin\chi. \label{elliptic}%
\end{equation}

This is the elliptic sine-Gordon equation. It is well investigated now
\cite{elliptic}. Screw dislocations in the elasticity theory, magnetization
distribution in easy plane magnets, Berezinskii-Kosterlitz-Thouless transition
and other systems can be described by this equation. It has a rich variety of
solutions including arrays of vortices in the London limit. Vortex solutions
are determined at the plane, punctured in the vicinity of the vortex cores.
The N-vortex solution for arbitary positions of vortices can be written in the
extreme London limit $\delta\rightarrow\infty$ as follows%
\begin{equation}
\chi=\sum_{k}^{N}n_{k}\arctan\dfrac{x-x_{k}}{y-y_{k}}. \label{linear}%
\end{equation}

\bigskip

Solutions of the nonlinear equation (\ref{elliptic}) was found, for instance,
in \cite{borisovtankeevshatalov}. A structure of the solitary vortex can be
analyzed if we note that the radial equation
\begin{equation}
\frac{d^{2}\chi}{dr^{2}}+\dfrac{1}{r}\frac{d\chi}{dr}=\frac{1}{\delta^{2}}%
\sin\chi\label{radial}%
\end{equation}
belongs to the third Painlev\'{e} class \cite{newel}, This equation was
thoroughly investigated in \cite{novokshenov}. Its solutions are shown to have
logarithmic asymptotics at $r\rightarrow0$ and oscillate at $r\rightarrow
\infty$.

In conclusion, we have considered an effect of the crystal lattice symmetry
and boundness of the kinetic term operator spectrum on electrodynamic
properties of the strong coupling superconductor.

\begin{acknowledgements}
I would like to thank E. Kudinov, Yu. Kuzmin and B. Shalaev for
discussions. The work was supported by the Russian Foundation for
Basic Research, grant No 02-02-17667.
\end{acknowledgements}
%

\begin{figure}
[b]
\begin{center}
\includegraphics[
natheight=5.021500cm,
natwidth=7.378500cm,
height=5.0215cm,
width=7.3785cm
]%
{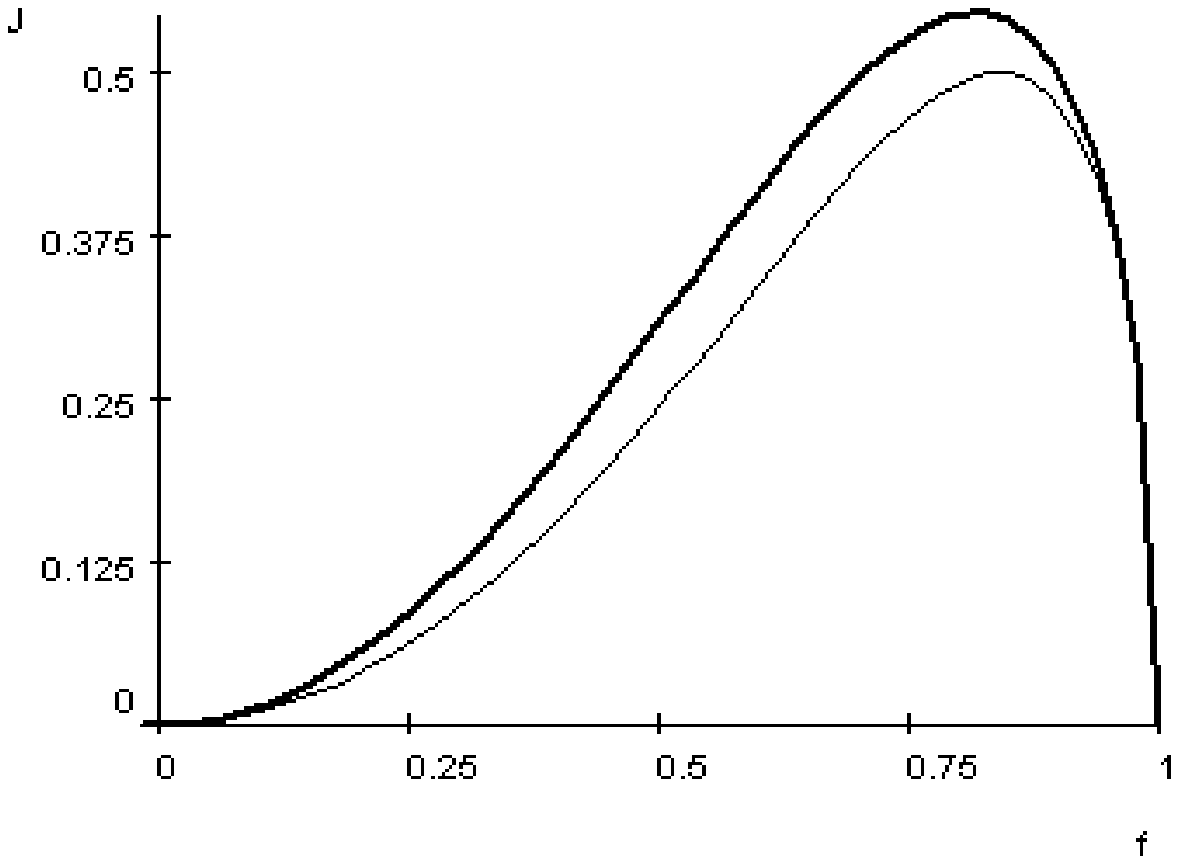}%
\caption{ Current -- order parameter relation for the thin film. Thick line --
the continuum limit; thin line -- the narrow band case with $k=1$. The maximum
position gives the critical current value. }%
\label{fig1}%
\end{center}
\end{figure}
%

\begin{figure}
[t]
\begin{center}
\includegraphics[
natheight=4.817200cm,
natwidth=7.125900cm,
height=4.8172cm,
width=7.1259cm
]%
{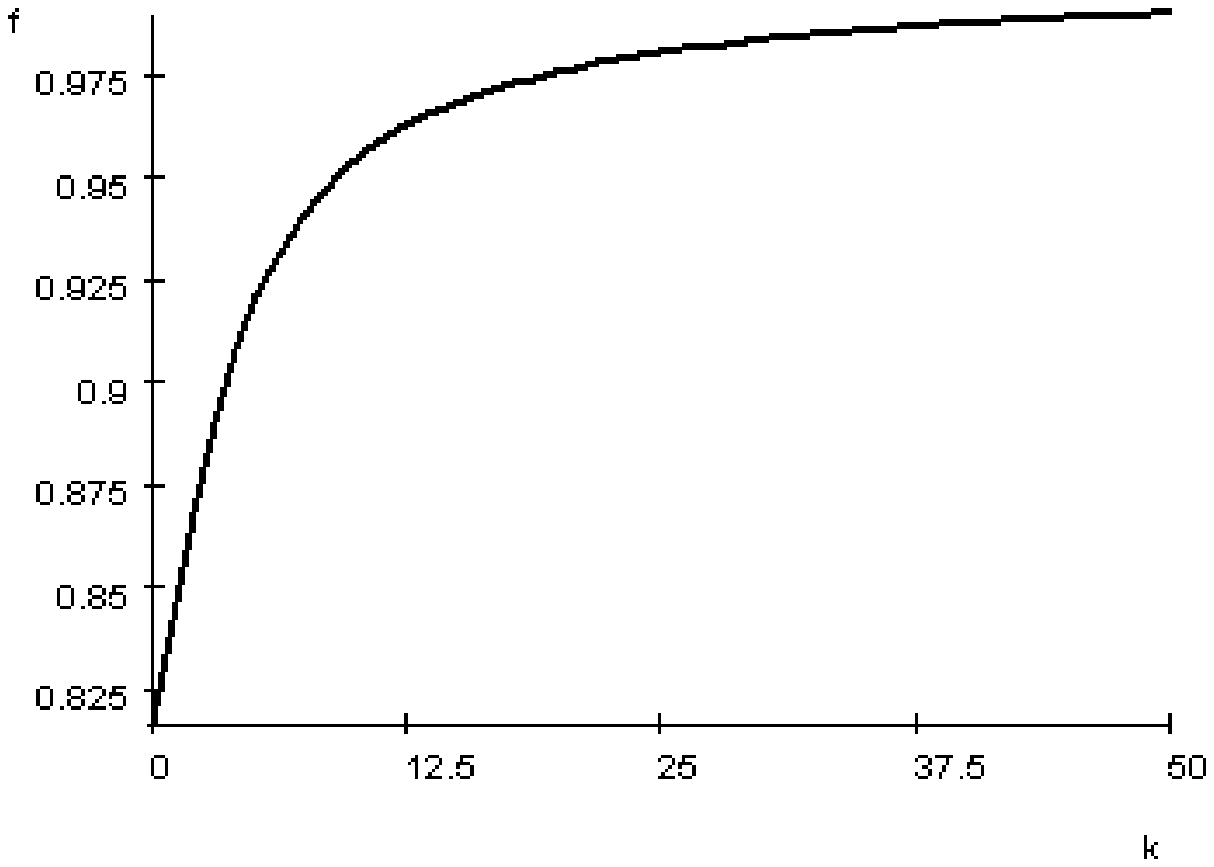}%
\caption{Current maximum position as a function of $k.$}%
\label{fig2}%
\end{center}
\end{figure}
%

\begin{figure}
[b]
\begin{center}
\includegraphics[
natheight=2.157700in,
natwidth=2.780400in,
height=2.1577in,
width=2.7804in
]%
{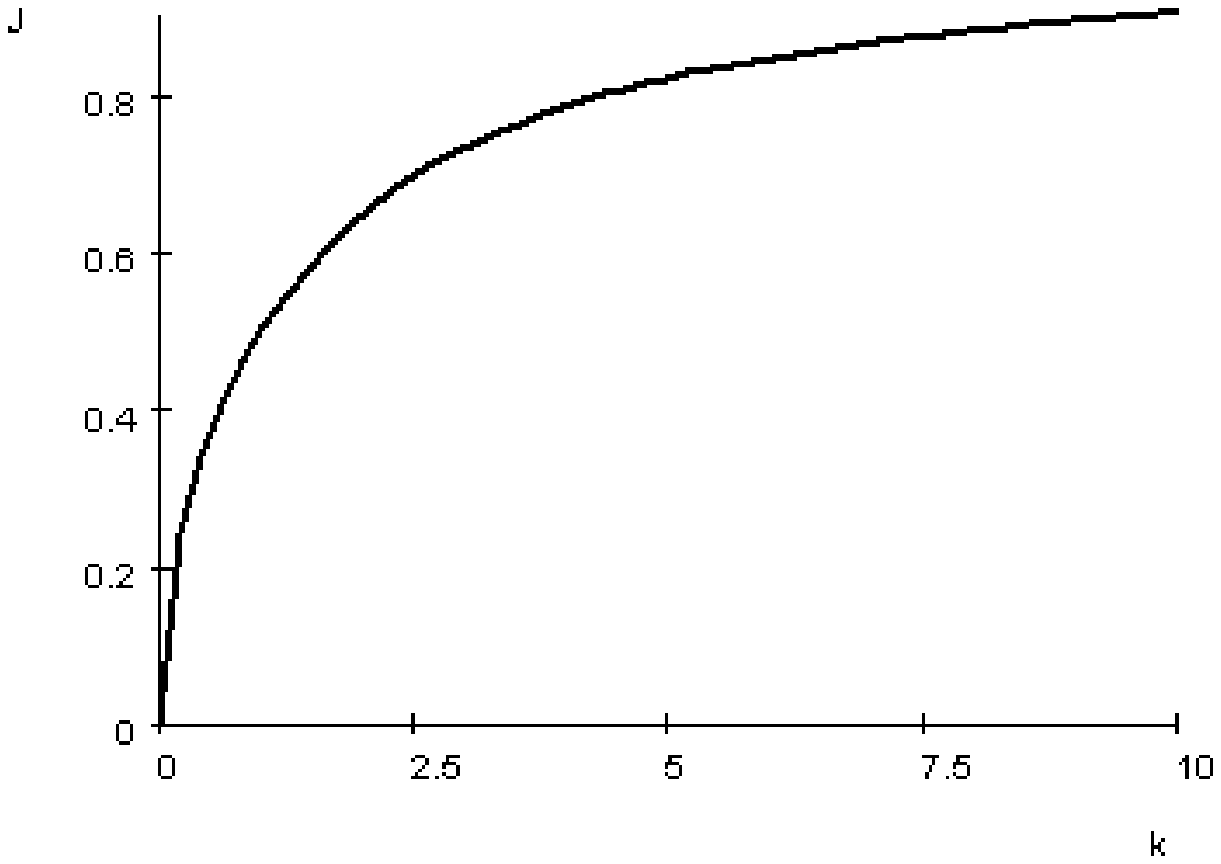}%
\caption{ Maximum dimensionless current $J$ as a function of $k.$}%
\label{fig3}%
\end{center}
\end{figure}


\begin{thebibliography}{99}                                                                                               %


\bibitem {brezin}{E. Brezin, D.R. Nelson, and A. Thiaville, Phys. Rev. B
\textbf{31}, \textit{11}, 7124-7132 (1985)}.

\bibitem {We}{S.A. Ktitorov, B.N. Shalaev, and L. Jastrabik, Phys. Rev. B
\textbf{49}, \textit{21}, 15248-15252 (1994)}.

\bibitem {degennes}{P.G. de Gennes, \textit{Superconductivity of metals and
alloys}, (W.A. Benjamin, Inc., New York, Amsterdam, 1966)}.

\bibitem {kud}{E.K. Kudinov, Solid State Physics (St. Petersburg),
\textbf{44}, 667 (2002)}.

\bibitem {sherst}{S.A. Ktitorov, V.S. Sherstinov, Jastrabik, and L. Soukup in
\textit{Weak Superconductivity}, ed. S. Benachka, P. Seidel, and V. Strib,
Bratislava, 1994, pp. 300-305}.

\bibitem {soliton}{V.E. Zaharov, S.V. Manakov, S.P. Novikov and L.P.
Pitaevskij,\textit{Soliton Theory: the Inverse Scattering Method}, (Moscow,
Mir, 1980). }

\bibitem {elliptic}{G. Leibbrandt, Phys. Rev. B \textbf{15}, \textit{7}, 3353
(1977)}.

\bibitem {borisovtankeevshatalov}{A.B. Borisov, A.P. Tankeev and A.G.
Shagalov, Metal Physics (Ekaterinburg) \textbf{60}, 467 (1985)}.

\bibitem {newel}{A. Newell, \textit{Solitons in Mathematics and Physics},
Society for Industrial and Applied Mathematics, 1985}.

\bibitem {novokshenov}{V.Yu. Novokshenov, A.G. Shagalov, Theor. Math. Phys
(Moscow), \textbf{111}, 15 (1997)}.
\end{thebibliography}
\end{document}